\documentclass[prl,twocolumn,amsmath,amssymb,groupedaddress,superscriptaddress]{revtex4}
\usepackage{graphics}

\def\be{\begin{equation}}
  \def\ee{\end{equation}}

\begin{document}

\title{Bose--Einstein Condensation and strong--correlation behavior of phonons in ion traps}

\author{D. \surname{Porras}}
\email{Diego.Porras@mpq.mpg.de} \affiliation{Max-Planck-Institut
f\"ur Quantenoptik, Hans-Kopfermann-Str. 1, Garching, D-85748,
Germany.}
\author{J.~I. \surname{Cirac}}
\email{Ignacio.Cirac@mpq.mpg.de}
\affiliation{Max-Planck-Institut f\"ur Quantenoptik, Hans-Kopfermann-Str.
1, Garching, D-85748, Germany.}

\begin{abstract}
We show that the dynamics of phonons in a set of trapped ions
interacting with lasers is described by a Bose--Hubbard model
whose parameters can be externally adjusted. We investigate the
possibility of observing several quantum many--body phenomena,
including (quasi) Bose--Einstein condensation as well as a
superfluid--Mott insulator quantum phase transition.
\end{abstract}

\date{\today}

\maketitle

Systems of ultracold bosons present a rich variety of fascinating
phenomena, like Bose--Einstein Condensation (BEC)
\cite{originalEinsteinBose}, or the superfluid--Mott insulator
(SI) quantum phase transition \cite{Fisher}. At present, there
exist very few physical systems in which these effects can be
observed. Atomic gases at ultra low temperature constitute a
unique system in this context, since their physical parameters can
be adjusted using external fields, something which has enabled the
observation of BEC \cite{AndersonDavis} or the SI
transitions \cite{MottBloch}, for example. In this paper we show
that phonons in a crystal of trapped ions interacting with lasers
provides us with another system where all these phenomena can be
observed in a very clean way. As in the case of neutral atoms, the
physical parameters describing the phonon dynamics can be adjusted
using lasers. Furthermore, individual addressing yields novel
possibilities for investigating novel physical situations.

In the set up we consider here, phonons are associated to the
motion of the trapped ions. Coulomb interaction induces the
transmission of phonons from one ion to another, whereas
anharmonicities in the trapping potentials give rise to an
effective phonon--phonon interaction. Thus, an ion crystal is
analogous to an optical lattice \cite{Jaksch},
whereby the ions play the role of lattice sites and the phonons
that of the atoms. An important feature of ion crystals is that,
due to energy conservation, phonons cannot be created or
annihilated, i.e. the phonon number is conserved. This is in
contrast with usual solid state systems, where phonons are
subjected to processes that do not conserve their number thus
preventing them from reaching, e.g., BEC. Furthermore, the
theoretical development
\cite{CiracZoller95, AMOReview} and
experimental progress \cite{Meekhof,experIons,LeibfriedReview}
experienced by the field of trapped ion quantum computation during
the last years can be exploited in the present context to gain
access to physical observables which are not reachable in other
systems.

Let us consider a set of $N$ trapped ions confined by external
electric potentials and which move around their equilibrium
positions. The Hamiltonian describing this situation can be
written as $H=K + V_0 + V_{\rm Coul}$, where $K$ describes the
kinetic energy, $V_0$ the trapping potential, and $V_{\rm Coul}$ the
Coulomb interaction between the ions. We will assume that: (i) the
motion of the ions along one particular direction, say
$\textbf{x}$, is decoupled from the motion along the other
directions; (ii) the trapping potential along $\textbf{x}$ for
each ion is practically harmonic with frequency $\omega$, i.e. it
is given by $\case{1}{2} m \omega^2 x_i^2$, where $x_i$ denotes
the operator corresponding to the displacement of the $i$--th ion
from its equilibrium position; (iii) the displacements around the
equilibrium position are much smaller than the distances between
ions; (iv) the Coulomb energy is small compared to the potential
energy, i.e. $\beta:=e^2/(d_0^3 m \omega^2)\ll 1$, 
where $d_0$
denotes the average separation between ions. The first requirement
(i) allows us to ignore the motional state of the ions along the
$\textbf{y}$ and $\textbf{z}$ directions. Condition (ii) allows us
to associate phonons to each of the ions in the usual way
\cite{AMOReview}: if the vibrational state of the $i$--th ion is given by
the $n$--th (Fock) excitation state of the corresponding harmonic
potential we will say that the ion has $n$ phonons and denote the
corresponding state by $|n\rangle$. The position operator of the
ion can be then written in terms of creation and annihilation
operators for the phonons, i.e. $x_i \propto (a_i+a_i^\dagger)$.
Condition (iii) allows us to express the Coulomb interaction between
ions $i$ and $j$ as $(e^2/d_{i,j}^3) x_i x_j$, where $d_{i,j}$ is
the distance between the ions. It is clear that this term will
induce hopping from phonons betweeen the ions since it contains
terms of the form $a_i^\dagger a_j$. On the other hand, condition
(iv) imposes that the phonon number is conserved, since the terms
of the form $a_i a_j$ would decrease the energy by $2\omega$,
something which cannot be compensated by the Coulomb interaction
which effectively switches off this process. Finally, the
anharmonicities of the trapping potential will be, in lowest
order, described by terms of the form $x_i^3$ or $x_i^4$. Again,
only energy--conserving terms will be important and thus only
those proportional to $a_i^\dagger a_i$ or $a_i^{\dagger 2} a_i^2$
will survive. The first one will add some small correction to the
trapping frequency,
whereas the second term can be associated to an effective
phonon--phonon interaction.

Thus, we have shown that the dynamics of the phonons in an ion
crystal will contain hopping terms between different ions as well
as on--site phonon--phonon interactions, and therefore they will
be described by a Bose--Hubbard model (BHM). Typically, the trapping
anharmonicities will be very small. However, they can be enhanced
by using off--resonant lasers. For instance, one may induce
repulsive phonon--phonon interactions by placing the
ions near the maximum of a standing wave, something which will
induce an AC--Stark shift proportional to $\cos(k x_i)^2 \approx 1
- (k x_i)^2 + (1/3) (k x_i)^4$, where $k$ is the wave-vector of
the laser. 
By placing them at the mininum, one obtains an
attractive phonon--phonon interaction.

There are several physical set--ups which realize a BHM as explained above. In the following we will concentrate in
the simplest one, which consists of ions in a linear trap and
which gives rise to a 1D system. Let us emphasize, however,
that with ions in microtraps \cite{deVoe} or in Penning traps \cite{Penning} one can realize
higher dimensional situations.

In a linear trap, ions are arranged in a Coulomb chain. Phonons
moving along the chain cannot be used in the way we described
above since for them $\beta \agt 1$ \cite{Dubin,Steane}.
However, transverse phonons
corresponding to the radial modes fulfill $\beta \ll 1$ and thus
are perfectly suited for our purposes. The radial phonon dispersion
relation has a large gap of the order of $\omega$, giving rise to
the phonon--number conservation, and has a bandwidth of the order
of $\beta\omega$. Therefore, we take $\textbf{x}$ as one of the
transverse directions and $\textbf{z}$ the trap axis. The
Hamiltonian that describes the motion in a chain with $N$ ions is given
by:
\begin{eqnarray}
V_0 &=& \frac{1}{2} m  \sum_{i=1}^N \left( \omega_{x}^2 x_i^2 +
\omega_y^2 y_i^2 +\omega_z^2 z_i^2 \right)  \\
V_{\rm Coul} &=& \sum_{i>j}^N \frac{e^2}{\sqrt{(z_i -z_j)^2 + (x_i
- x_j)^2 + (y_i - y_j)^2}}. \nonumber
\label{ion.chain}
\end{eqnarray}
where $\omega_{\alpha}$, $\alpha$ = $x,y,z$, are the trapping frequencies in 
each direction, and we define $\beta_{\alpha}$ as the corresponding ratios between
Coulomb and trapping energy.
If $\omega_{x,y} \gg \omega_z$ the ions form a chain along the ${\bf z}$ axis
and occupy equilibrium positions $z_i^0$. Phonons in the {\bf x} direction can be 
described approximately by:

\begin{eqnarray}
H_{x0} &=& \sum^N_{i=1} (\omega_x + \omega_{x,i}) a^{\dagger}_i a_i
+ \sum^N_{j>i} t_{i,j} (a^{\dagger}_i a_j + a_i a^{\dagger}_j )
 \\
\omega_{x,i} &=& 
- \frac{1}{2} \sum^N_{j' \neq i} 
\frac{e^2/(m \omega_x^2)}{|z_i^0 \! - \! z_{j'}^0|^3} \ \omega_x, \ \
t_{i,j} = \frac{1}{2} \frac{e^2/(m \omega_x^2)}{|z_i^0 \! - \! z_{j'}^0|^3} \ \omega_x .
\nonumber
\label{tight.binding}
\end{eqnarray}
$\omega_{x,i}$, $t_{i,j}$ are spatial dependent shifts of the trapping frequency, and effective hopping energies, respectively. Note that both of them are of the order of $\beta_x \omega_x$. The approximations that lead to (\ref{tight.binding}) are: (i) In $V_{Coul}$, we keep only second order terms in the displacements of the ions around the equilibrium positions. Higher order terms are of the form $x^4$, $x^2 y^2$, $z x^2$. They can be neglected assuming that $x_0/d_0, z_0/d_0 \ll 1$, with $x_0$, $z_0$ the size of an individual ion's wave-packet. 
$x_0$ can be estimated by the size of the ground state in the radial trapping frequency and for typical parameters 
($d_0$ = 5 $\mu m$, $\omega_x$ = 10 MHz, also used below) 
we have $x_0/d_0$ $\approx$ $10^{-3}$. 
In the case of $z_0$, one has to consider the collective nature of the axial modes, because $\beta_z$ $\gg 1$ if $N \gg 1$. If we consider axial modes at a finite temperature, $z_0$ is given by the thermal fluctuations of the position of the ion. In the limit $N \gg 1$, we can estimate $z_0^2 \approx \frac{\hbar}{2 m \omega_z \sqrt{\beta_z \log{\beta_z}}} \frac{k_B T}{\hbar \omega_z}$, 
which means that $z_0/d_0$ $\approx$ $10^{-3} k_B T / \hbar \omega_z$, with $\omega_z$ = $100$ kHz, $N=100$ (see \cite{note}). (ii) We consider $\beta_x \ll 1$, and neglect the phonon number non--conserving terms in the couplings of the form $x_i x_j$ $\propto$ $(a^{\dagger}_i + a_i)(a^{\dagger}_j + a_j)$.

We include the effect of a standing wave in $H_x$, such that a repulsive phonon--phonon interaction is induced \cite{AMOReview}:
\begin{eqnarray}
H_{sw} = F \sum^N_{i=1} | 0 \rangle_i \langle 0 | \cos^2(k x_i) .
\label{standing.wave}
\end{eqnarray}
$| 0 \rangle_i $ is the internal ground state of the ions. In
the  following we will assume that ions stay always in $| 0 \rangle_i $, and expand the standing-wave in the Lamb-Dicke parameter,
$\eta = k x_0/\hbar \omega_x$:
\begin{eqnarray}
H_{s \! w} = F  \sum^N_{i=1}
( 1
 \! + \! \eta^2  (a_i \! + \! a^{\dagger}_i)^2
  + \! \frac{1}{3} \eta^4  (a_i \! + \! a^{\dagger}_i)^4 \! + \! {\cal O}(\eta^6) ) .
\label{lamb.dicke}
\end{eqnarray}
The fourth order contribution contains a Hubbard interaction,
$U {a^{\dagger}_i}^2 {a_i}^2$, with $U = 2 F
\eta^4$ (note that $U < 0$ if the ions are placed at the minimum). The other terms in Eq. (\ref{lamb.dicke}) are: (i) phonon conserving terms that just give
corrections to the trapping frequency; (ii) phonon non--conserving terms, that rotate with frequency $\omega_x$. 
The non--conserving contributions can be adiabatically eliminated
if $F \eta^2 / \omega_x \ll 1$. For example, in the case of the second order terms, $F \eta^2 (a_i^2 + {a^{\dagger}_i}^2)$, a perturbative calculation shows that they only give harmonic corrections of the form $((F \eta^2)^2/\omega_x) (-2 a_i^{\dagger} a_i -1) + {\cal O} (F \eta^2)^3/\omega_x^2$. Thus the contributions from non--conserving term, either give corrections to the trapping frequency, or can be neglected when compared to $U$.

The final Hamiltonian takes the form of a BHM:
\begin{equation}
H_x = 
H_{x0} + \sum^N_{i=1} U {a_i^{\dagger}}^2 a_i^2,
\label{bose.hubbard}
\end{equation}
where we include in $H_{x0}$ the corrections 
from the standing wave. Note that
as long as the number of phonons is conserved,
$\omega_x$ in $H_{x0}$ is a global chemical potential that
does not play any role in the description of the system.

\begin{figure}
  \resizebox{2.in}{!}{\includegraphics{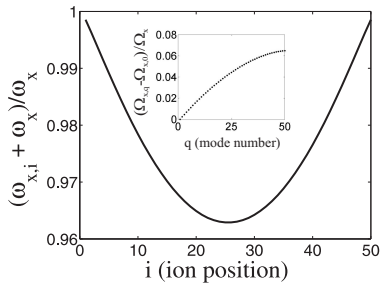}}
   \caption{Phonon trapping potential (\ref{tight.binding}) for the radial modes of a Coulomb
   chain with $N=$50 ions, and $\beta_x$ $=$ $10^{-2}$, as a function of the ion
   position along the chain. Inset: Spectrum of the
   radial collective modes, $\Omega_{x,q}$, that 
   diagonalize $H_{x0}$.}
\label{spectrap}
\end{figure}

\begin{figure}
  \resizebox{2.in}{!}{%
    \includegraphics{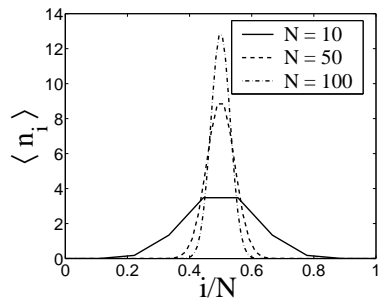}}
\caption{Mean phonon number $\langle n_i \rangle = \langle a^{\dagger}_i a_i \rangle$ 
along Coulomb chains with different number of ions, $N$,  
in the state with $N_{ph} = N$ phonons in the radial lowest mode. The width of the wavefunction in units
  of $N$, is $\propto 1/\sqrt{N}$, in
  accord with the scaling for the phonon trapping frequency, $\omega_c$ $\propto$ $1/N$.}
\label{wave}
\end{figure}

We discuss now the properties of the solutions of the
non-interacting Hamiltonian, $H_{x0}$. A quite unexpected result is that the Coulomb interaction induces the confinement of the radial phonons.
This is due to the fact that the distance between ions is larger
at the sides than at the center of the chain, and is well described by a quadratic dependence on the position of the ions. 
Thus, the corrections $\omega_{x,i}$ in Eq. (\ref{tight.binding}) are smaller for the ions placed at the center of the chain, in
such a way that the radial phonon field is confined (Fig. \ref{spectrap}).
The harmonic phonon confinement can be estimated by means of
Eq. (\ref{tight.binding}) in the limit $N \gg 1$. 
In this  case, the distance between ions at site $i$ satisfies \cite{Dubin}:
\begin{equation}
\frac{1}{(d(i')/d_0)^3} \approx \alpha - \gamma \left( \frac{i'}{N} \right)^2, \ \ \alpha = 3.4, \ \gamma = 18,
\label{distance.ions}
\end{equation}
where $i' = i - N/2$. One can use Eq. (\ref{distance.ions}) to
describe qualitatively the dependence  of $\omega_{x,i}$ with the
position. We include only Coulomb interaction between nearest--neighbors in order to
get analytical results. The spatial dependent part of the non--interacting
boson Hamiltonian is given, in this approximation, by:
\begin{eqnarray}
&& H_x /(\beta_x \omega_x)  = \\
&& \sum^N_{i=1} \frac{1}{2} \gamma ( \frac{i'}{N} )^2  \! a^{\dagger}_i a_i
+ \frac{1}{2} \left( \alpha - \gamma ( \frac{i'}{N} )^2 \right) (a^{\dagger}_i a_{i+1} + h.c.) .
\nonumber
\label{hamil.confinement}
\end{eqnarray}
In the limit of many ions and low energies, the continuum limit in this expression  describes a one dimensional system of bosons
trapped by the frequency $\omega_c \approx (8/N) \beta_x \omega_x$.
The lowest collective modes in the exact spectrum  show a linear dispersion that is well described by our estimation for $\omega_c$ (see inset of Fig. \ref{spectrap}).
Thus, we get the conclusion that radial phonons in ions in linear Paul traps are naturally confined by an approximate harmonic potential (see Fig. \ref{wave}).

\begin{figure}
  \resizebox{2.in}{!}{%
    \includegraphics{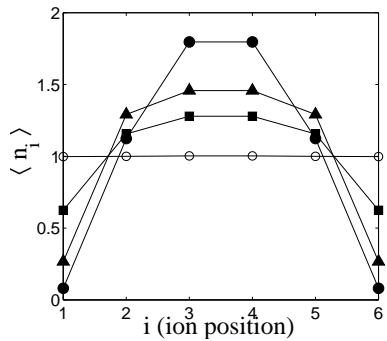}}
  \caption{Mean phonon number at each ion in the ground state of  a Coulomb chain with $N = 6$ ions, and $N_{ph} = 6$ phonons. $\beta_x = 0.01$, so that the nearest neighbor hopping terms are $t \approx 5 \ 10^{-3} \omega_x$. Black circles represent the phonon density without standing
  wave ($U = 0$), and show the confinement due to the phonon trapping potential. Empty circles: Mott
  phase when a standing wave is applied with $F \eta^2 = 0.1 \omega_x$, $\eta^2 = 0.1$,
  and $U  \approx 0.02 \omega_x > t$. Squares: $U = 0.01 \omega_x$, Triangles:
  $U = 5 \ 10^{-3} \omega_x$.}
\label{hubbard}
\end{figure}

Our ideas lead to the following two proposals of experiments with linear Paul traps:

{\em (i) Superfluid-Mott insulator transition and creation of a
superfluid  phonon state by adiabatic evolution.} Hamiltonian
(\ref{bose.hubbard}) describes a BHM with the
peculiarity that hopping terms are positive, with a range that
is longer than the usual nearest-neighbor hopping in optical
lattices. However we can understand the properties
of our system by means of the better known model with nearest-neighbor
hopping only \cite{Fisher,Sachdev}.
Let us consider $t = (1/2) \beta_x \hbar \omega_x$, the characteristic hopping energy.
If the total number of phonons $N_{ph}$ is commensurate with the number 
of ions $N$, then, for values $U \gg t$, the ground state of Hamiltonian (\ref{bose.hubbard}) is a Mott insulator, well described by a product of Fock states of $N_{ph}/N$
in each ion (note that phonon confinement, $\hbar \omega_c$ is also of order $t$, so that condition $U \gg t$ ensures a uniform phonon density).
On the other hand, the ground state for $U \ll t$ is a
superfluid with all the phonons in the lowest energy level. In
Fig. \ref{hubbard}, we present the results of an exact numerical
diagonalization of the complete phonon Hamiltonian (that is,
including also the phonon number non-conserving terms) for the case
$N_{ph}/N = 1$. The transition from the superfluid to the Mott
insulator, with one phonon per site, is evident in the evolution
of the phonon density as a function of the interaction $U$.

The properties of the BHM, allows us to propose an
experimental sequence that would lead to the observation of the SI
quantum phase transition: {\em (1)} The ion chain is cooled to the
state with zero radial phonons by laser cooling. {\em (2)}
Starting with a value $U \gg t$, the eigenstates of the system are
well described by Fock states localized at each ion. The ground
state of the phonon system can be created by means of
sequencies of blue/red-side band transitions, in a method that has been successfully implemented with single trapped ions (see
\cite{Meekhof}). {\em (3)} The value of $U$ is varied
adiabatically down to a given value $U_f$, in such a way that the
system remains in the ground state. At a given critical value $U_f
\approx t$, the system undergoes a transition to a phonon
superfluid. {\em (4)} The measurement of the ground state can be
accomplished by the coupling of the transverse phonons to a given
internal transition. One could apply, for example, a red sideband
pulse with intensity $g$, for a short time $t$. Under such
conditions, the probability of inducing a transition to the
excited internal state is $\propto \sum_n (\sin(\sqrt{n} g t))^2 P(n) \approx \sum_n
n g^2 t^2 P(n)$, where $P(n)$ is the probability of having $n$
phonons. This method would allow us to measure the mean phonon
number. By resolving individually the photoluminescence from each
ion, one could observe features of the BEC, or SI
transition in the variations of the phonon density along the
chain. One could also apply well known methods to determine
$P(n)$, or even the whole quantum tomography of the phonon quantum states \cite{Meekhof,LeibfriedReview}.

{\em (ii) (Quasi) Bose--Einstein Condensation by evaporative laser
cooling.}
We propose an experiment that is akin to the usual BEC of cold
atoms in harmonic traps. First, we note that techniques for
cooling of trapped ions, like laser cooling \cite{LeibfriedReview,cooling} can
only be used to destroy phonons. The existence of the trapping
phonon potential in ion traps allows us to propose the combination
of laser cooling with the idea of evaporative cooling. A
possible experimental sequence would be as follows: {\it (1)} Start
with a Coulomb chain after usual Doppler cooling, that is, a chain
with a given number of phonons per site, and induce a small
phonon--phonon interaction $U \ll t$, so that the system remains in
the weak interacting regime. {\it (2)} Apply laser cooling at the
sides of the Coulomb chain, in such a way that the higher energy
phonons on the top of the confinement potential are destroyed
(evaporated). {\it (3)} The interaction $U$ induces collisions
that thermalize the phonons to a lower temperature. Several cycles
of laser cooling / thermalization could be applied until the
system is cooled below the critical temperature. Detection of the
BEC could be accomplished along the same lines exposed above for
the case of the BHM. 
Note that in the case of Coulomb chains (1D) considered here, 
(quasi) BEC is possible in finite size systems only.

We have shown that phonons in a
system of trapped ions can be manipulated in such a way that they
undergo BEC, or a SI transition. The main ingredients of our
proposal are: {\it (1)} The fact that phonons can have a large
energy gap that suppresses processes that do no conserve the
number of phonons. {\it (2)} Phonon--phonon interactions
(anharmonicities) can be induced by placing the ions in a
standing--wave. {\it (3)} In the particular case of ions in a
linear trap, radial phonons would be suitable for this
proposal, with the advantage that the Coulomb interaction provides
us with an approximately harmonic phonon confinement.

In this work we have exposed only a few applications of this idea,
but phonons  in trapped ions could be used to study quantum phases
with a degree of controllability that is not possible with cold
neutral atoms. Individual addressing would allow us to design
Hubbard Hamiltonians with local interactions that change at will
from site to site. Different directions of the radial modes, or
different internal states of the ions could play the role of effective
spins \cite{porras} for the phonons.
On the other hand, one could also reach the regime dominated by the
repulsive interaction and create, thus, a Tonks-Girardeau gas of phonons \cite{girardeau}
in a Coulomb chain (this idea was implemented recently with
optical lattices \cite{belen}). In a
very promising approach, 2D systems of arrays of microtraps
\cite{deVoe}, or ions in Penning traps \cite{Penning}, could be considered, because phonons
transverse to the crystal plane satisfy the conditions required by our proposal.

Work supported by the European projects CONQUEST, RESQ and TOPQIP, and
the Kompetenz\-netz\-werk Quanten\-informations\-verarbeitung der
Bayerischen Staatsregierung.

\newpage


\begin{thebibliography}{99}
\bibitem{originalEinsteinBose} A. Einstein, Sitzber. Kgl. Preuss. Akad. Wiss., 261 (1924); 3 (1925); S.N. Bose, Z. Phys. {\bf 26}, 178 (1924). 
\bibitem{Fisher} M.P.A. Fisher {\it et al.}, Phys. Rev. B {\bf 40}, 546 (1989).
%
\bibitem{AndersonDavis} M.H. Anderson {\it et al.}
Science {\bf 269}, 198 (1995); K.B. Davis {\it et al.},
Phys. Rev. Lett. {\bf 75}, 3969 (1995).
%
\bibitem{MottBloch} M. Greiner {\it et al.},
Nature (London) {\bf 415}, 39 (2001).
%
\bibitem {Jaksch} D. Jaksch {\it et al.},
Phys. Rev. Lett. \textbf{81}, 3108 (1998).
%
\bibitem{CiracZoller95} J.I. Cirac and P. Zoller, Phys. Rev. Lett. \textbf{74}, 4091 (1995).

\bibitem{AMOReview} J.I. Cirac {\it et al.}, Adv. At. Mol. Opt. Phys. 37, 237 (1996).
%
\bibitem{Meekhof} D.M. Meekhof {\it et al.},
Phys. Rev. Lett. {\bf 76}, 1796 (1996).
%
\bibitem{experIons} D.J. Wineland {\it et al.},  J. Res. NIST {\bf 103}, 259 (1998);
D. Leibfried {\it et al.}, Phys. Rev. Lett. {\bf 89}, 247901 (2002);
S. Gulde {\it et al.}, Nature (London) {\bf 421}, 48 (2003); F. Schmidt-Kaler {\it et al.}, Nature (London) {\bf 422}, 408 (2003).
%
\bibitem{LeibfriedReview} D. Leibfried {\it et al.}, 
Rev. Mod. Phys. \textbf{75}, 281 (2003).
%
\bibitem{deVoe} Ralph G. de Voe, Phys. Rev. A \textbf{65}, 063407 (2002) .
%
\bibitem{Penning} W.M. Itano {\it et al.},
Science {\bf 279}, 686 (1998); T.B. Mitchell {\it et al.},
Science {\bf 282}, 1290 (1998).

\bibitem{note} Note that higher order contributions in the expansion of the Coulomb energy of the form $x^2 y^2$, $x^2 z$, contain terms that do not conserve the number of phonons in the x-direction. However, these terms can be eliminated in a rotating wave approximation under the condition that the trapping frequencies in each direction are different: $\omega_x \neq \omega_y \neq \omega_z$.

\bibitem{Dubin} D.H.E. Dubin and T.M. O'Neil, Rev. Mod. Phys. {\bf 71}, 87 (1999).
\bibitem{Steane} A. Steane, Appl. Phys. B: Lasers Opt. \textbf{64}, 623 (1997); D.F.V. James, Appl. Phys. B: Lasers Opt. \textbf{66}, 181 (1998).


%
\bibitem{Sachdev} S. Sachdev, {\it Quantum phase transitions}, Cambridge University Press (1999).
%
%
\bibitem{cooling} C. Monroe {\it et al.},
Phys. Rev. Lett. {\bf 75}, 4011 (1995); Ch. Roos {\it et al.}
Phys. Rev. Lett. {\bf 83}, 4713 (1999).
%
\bibitem{porras} D. Porras and J.I. Cirac, Phys. Rev. Lett. {\bf 92}, 207901 (2004).
\bibitem{girardeau} M. Girardeau, J. Math. Phys. {\bf 1}, 516 (1960).
\bibitem{belen} B. Paredes {\it et al.}, Nature (London) {\bf 429} 277.
%
\end{thebibliography}
\end{document}